\documentclass[]{spie}  

\usepackage{amsmath,amsfonts,amssymb}
\usepackage{graphicx}
\usepackage{caption}
\usepackage{subcaption}

\usepackage[colorlinks=true, allcolors=blue]{hyperref}

\title{Utilizing U-Net Architectures with Auxiliary Information for Scatter Correction in CBCT Across Different Field-of-View Settings}

\author[a,b]{Harshit Agrawal}
\author[b]{Ari Hietanen}
\author[a]{Simo S\"{a}rkk\"{a}}
\affil[a]{Aalto University, Espoo, Finand.}
\affil[b]{Planmeca Oy., Helsinki, Finland.}

\authorinfo{Further author information: (Send correspondence to Harshit Agrawal.)\\E-mail: harshit.agrawal@aalto.fi \\ Accepted for publication in Proc. SPIE, Medical Imaging 2024: Physics of Medical Imaging}

\begin{document} 
\maketitle
\begin{abstract}
Cone-beam computed tomography (CBCT) has become a vital imaging technique in various medical fields but scatter artifacts are a major limitation in CBCT scanning. This challenge is exacerbated by the use of large flat panel 2D detectors. The scatter-to-primary ratio increases significantly with the increase in the size of FOV being scanned. Several deep learning methods, particularly U-Net architectures, have shown promising capabilities in estimating the scatter directly from the CBCT projections. However, the influence of varying FOV sizes on these deep learning models remains unexplored. Having a single neural network for the scatter estimation of varying FOV projections can be of significant importance towards real clinical applications. This study aims to train and evaluate the performance of a U-Net network on a simulated dataset with varying FOV sizes. We further propose a new method (Aux-Net) by providing auxiliary information, such as FOV size, to the U-Net encoder. We validate our method on 30 different FOV sizes and compare it with the U-Net. Our study demonstrates that providing auxiliary information to the network enhances the generalization capability of the U-Net. Our findings suggest that this novel approach outperforms the baseline U-Net, offering a significant step towards practical application in real clinical settings where CBCT systems are employed to scan a wide range of FOVs.
\end{abstract}

\section{INTRODUCTION}
\label{sec:intro}  

One major drawback of cone-beam computed tomography (CBCT) systems is the increase in scatter-to-primary signal ratio due to the use of large flat panel 2D detector and the proximity of detector to the patient.
Several deep learning-based methods have been proposed in recent years for the scatter estimation. A U-Net architecture was proposed for low-frequency scatter estimation for different tube voltages, anatomies, and noise levels \cite{maier2019}. This architecture estimated scatter signal directly from the scatter-affected projections. Similarly, other versions of U-Net and its variants have been investigated for the scatter estimation \cite{nomura2019projection, lee2019deep, roser2021x}. 

In CBCT systems, the scatter may vary significantly with the size of scanning field-of-view (FOV). The scatter-to-primary ratio increases 4.9-fold for a 17x12 cm FOV in comparison to a 6x6 cm FOV \cite{pauwels2021scatter}. In modern CBCT systems, the FOV size can be even larger, which will increase the scatter-to-primary ratio even more. While the applicability of deep learning networks has been demonstrated for estimating the scatter for a fixed FOV, no study has been done to evaluate deep learning network's performance with changes in FOV size. Moreover, it remains unclear if a single neural network can generalize to different FOVs. In practice, having a single network for multiple FOVs is important as CBCT systems are employed to scan a wide range of FOVs in real clinical settings.

In this study, we train and evaluate a single U-Net for the scatter estimation performance \cite{maier2019} on varying FOV sizes. Further, we propose a simple method to improve the performance of the single network for multiple FOV sizes by providing auxiliary FOV size information to the encoder (Aux-Net), which outperforms the baseline U-Net.

\section{Materials and Methods}
\subsection{Dataset}
We utilized two different sources to generate the training dataset. We used 3 CT reconstructions from  HNSCC-3DCT-RT dataset \cite{bejarano2018head, clark2013cancer} for generating training data. Further, 3 reconstructions were acquired by scanning 3 anthropomorphic phantoms using GE CT scanner, out of which we used 2 CT reconstructions for training and 1 for testing. We used Monte Carlo based X-ray simulations \cite{badal2009accelerating, agrawal2023deep} to generate primary (without scatter) and the corresponding scatter projections. For each reconstruction in training, we simulated total 18 different sizes of FOVs. For test phantom, we simulated a total of 30 FOV sizes which were different than the FOV sizes used in the training. The FOVs used in the training and the testing are given in the table \ref{tab:FOVs}. We simulated the  geometry of Planmeca Viso G7{\textregistered} CBCT scanner (Planmeca Oy, Helsinki, Finland) for the jaw protocol. The source-to-detector distance was 700 mm and the source-to-object distance was 490 mm. For each phantom in the training, the projections were simulated from 0 to 210 degrees with an spacing of 2.1 degrees. We simulated 100 projections for each phantom. We further augmented our dataset by simulating each projection 10 times with a different random seed to generate random direction and energy of X-ray. We averaged these simulated projections with each other to obtain 10 different noise levels for training. For example, the average projection of two simulated projections has different noise level than the average of 10 simulated projections. For test dataset, we simulated 500 views for each phantom ranging from 0 to 210 degrees. We also scanned an anthropological phantom using Planmeca Viso G7{\textregistered} CBCT scanner (Planmeca Oy, Helsinki, Finland) for $170 \times 170$ mm FOV to compare the results of the scatter correction methods qualitatively.

\begin{table}[ht]
\caption{The FOV sizes used for the training (X) and the testing (\checkmark).}
\centering   
\begin{tabular}{|c|*{17}{c|}}
\hline
diameter (mm) & \multicolumn{16}{c|}{height (mm)} \\
\hline
& 30 & 40 & 50 & 60 & 70 & 80 & 90 & 100 & 110 & 120 & 130 & 140 & 150 & 160 & 170 & 180 \\
\hline
120 & X & & & X & & & X & & & X & & & X & & & X\\
\hline
130 & & \checkmark & \checkmark & & \checkmark & \checkmark & & \checkmark & \checkmark & & \checkmark & \checkmark & & \checkmark & \checkmark & \\
\hline
140 & X & & & X & & & X & & & X & & & X & & & X\\
\hline
150& & \checkmark & \checkmark & & \checkmark & \checkmark & & \checkmark & \checkmark & & \checkmark & \checkmark & & \checkmark & \checkmark & \\
\hline
160 & X & & & X & & & X & & & X & & & X & & & X\\
\hline
170 & & \checkmark & \checkmark & & \checkmark & \checkmark & & \checkmark & \checkmark & & \checkmark & \checkmark & & \checkmark & \checkmark & \\
\hline
\end{tabular}
\label{tab:FOVs}
\end{table}

\subsection{Network Architecture}
\begin{figure}[htbp]
\includegraphics[width=\textwidth]{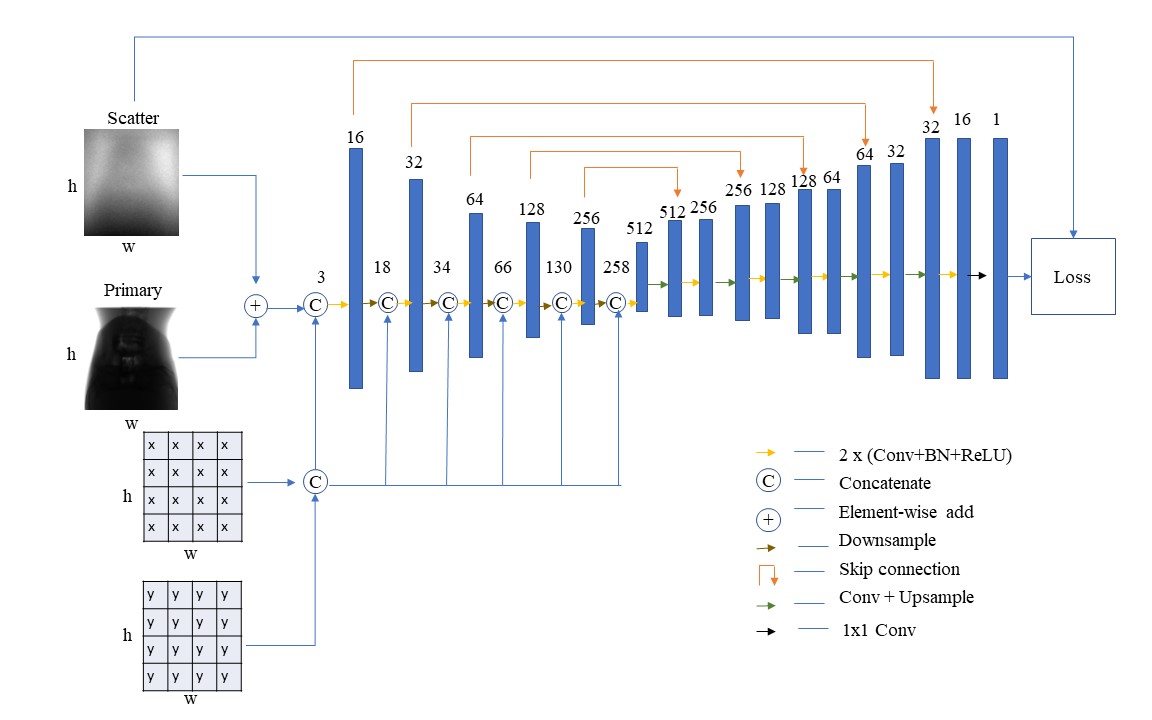}
\caption{Aux-Net architecture. $x$ in the first auxiliary channel is width $w$ normalized by maximum width of detector (250 mm) and  $y$ in the second auxiliary channel is height $h$ normalized by maximum height of detector (300 mm). The auxiliary channels are recreated at each level of the encoder to match the dimensions of the corresponding feature maps.}
\label{fig:auxnet}
\end{figure}

The architecture of the proposed Aux-Net is shown in the Fig.\ref{fig:auxnet}. To incorporate auxiliary information about the FOV shape of the CBCT projections, we create two additional channels, both having the same dimensions as the input features. The first channel contains width information and the second channel contains height information. The height and width are divided by the respective detector maximum sizes, i.e. 300 mm and 250 mm, respectively. Since magnification factor is same for all simulation geometries, the FOV shape is directly related with the shape of the projection. These two channels are concatenated to each level's input features in the encoder. For each level of the encoder, the channels are sized exactly to the input features to that level. Aux-Net has a slightly increased 7.275697 million parameters compared to the baseline U-Net's 7.257552 million trainable parameters. The baseline U-Net has same architecture as Aux-Net except the presence of auxiliary channels.
\subsection{Training and Evaluation}
For the training, the simulated primary and scatter projections were added and divided by the simulated flats. The input to the model was further linearized. All the input and target were resized to $300\times250$. The model was trained with a combination of mean-square-error (MSE) and high-frequency loss to penalize the high-frequency component in the predicted scatter.
For the evaluation of neural network predictions, the mean-absolute-percentage-error (MAPE) \cite{maier2019} and MSE were calculated between the upsampled neural network scatter estimates and Monte Carlo simulated ground truth scatter.
\section{Results}

\subsection{Scatter estimation on simulated data}
\begin{figure}[htbp]
    \centering
    \begin{subfigure}[b]{\textwidth}
        \includegraphics[width=\textwidth]{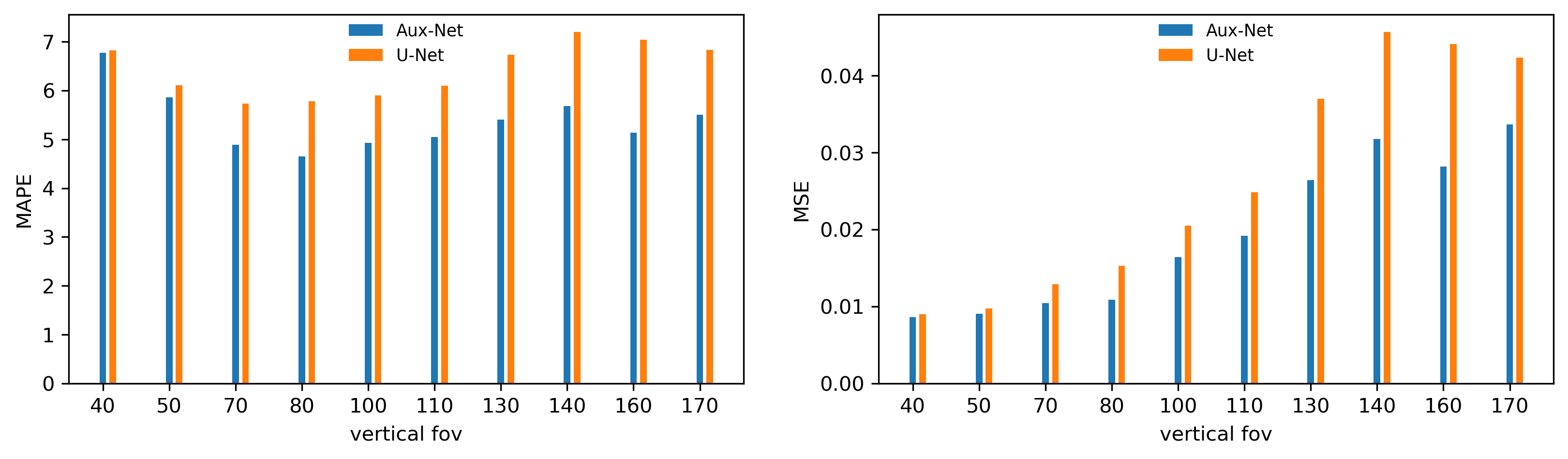}
        \caption{MAPE and MSE errors for U-Net and AuxNet for FOV diameter of 130 mm. X-axis shows the FOV heights in mm.}
        \label{fig:subfig1}
    \end{subfigure}
   
    \begin{subfigure}[b]{\textwidth}
        \includegraphics[width=\textwidth, ]{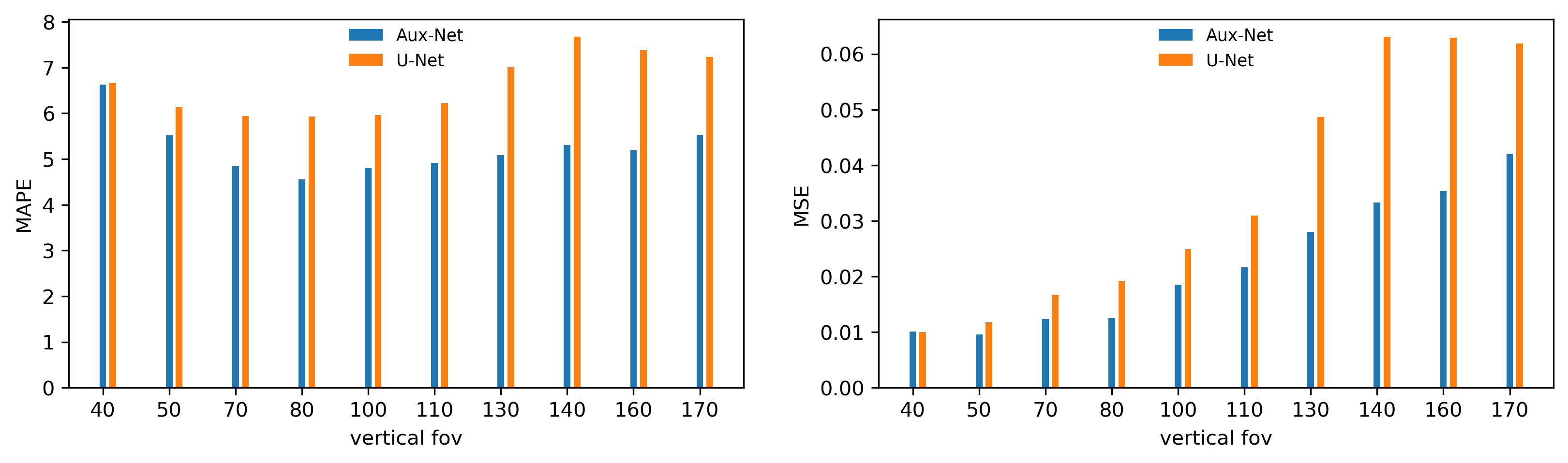}
        \caption{MAPE and MSE errors for U-Net and AuxNet for FOV diameter of 150 mm. X-axis shows the FOV heights in mm.}
        \label{fig:subfig2}
    \end{subfigure}

    \begin{subfigure}[b]{\textwidth}
        \includegraphics[width=\textwidth]{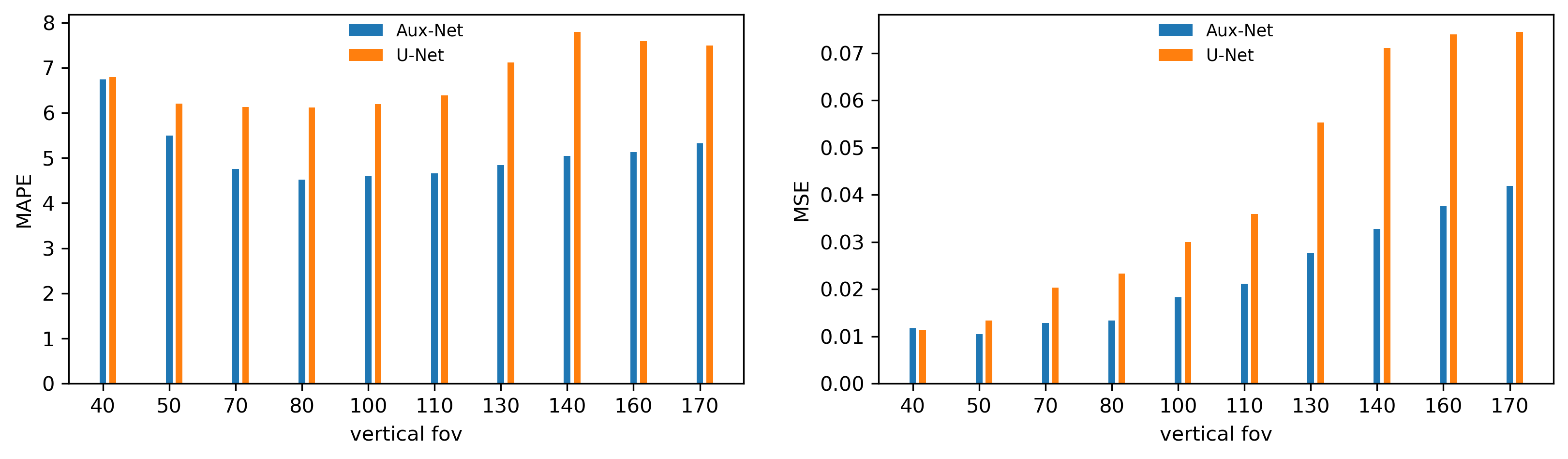}
        \caption{MAPE and MSE errors for U-Net and AuxNet for FOV diameter of 170 mm. X-axis shows the FOV heights in mm.}
        \label{fig:subfig3}
    \end{subfigure}\\

    \caption{Comparison of the scatter estimates from U-Net and AuxNet with Monte Carlo simulated ground truth scatter.}
    \label{fig:metrics}
\end{figure}

We compare our method Aux-Net with baseline U-Net and Monte Carlo simulated ground truth scatter projections. Fig. \ref{fig:metrics} shows MAPE and MSE errors between the ground truth and the neural network-based scatter estimates. The three figures show that proposed Aux-Net outperforms the baseline U-Net on all FOV sizes.

\subsection{Scatter estimation on phantom scan}
\begin{figure}[htbp]
    \centering
    \begin{subfigure}[b]{0.49\textwidth}
    \centering
        \includegraphics[height=3cm]{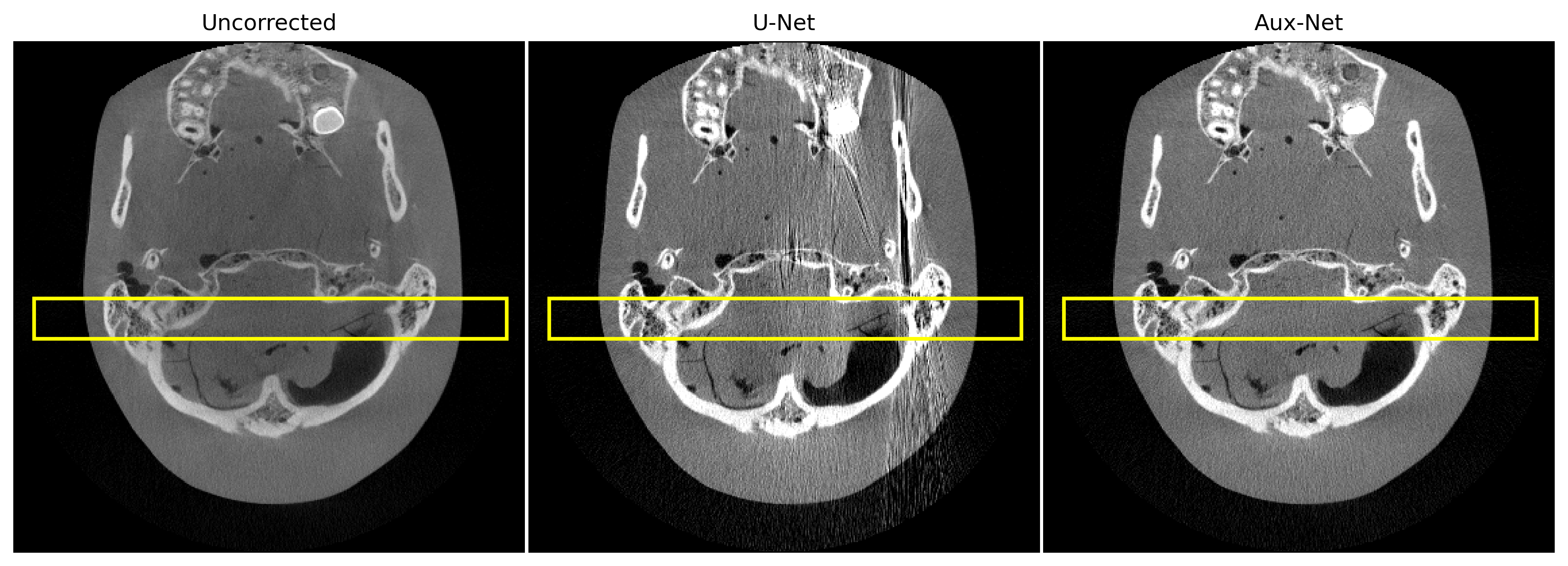}
        \caption{Axial image slice number 113.}
        \label{fig:subfig10}
    \end{subfigure}
   \hfill
    \begin{subfigure}[b]{0.49\textwidth}
        \centering
        \includegraphics[height=3cm]{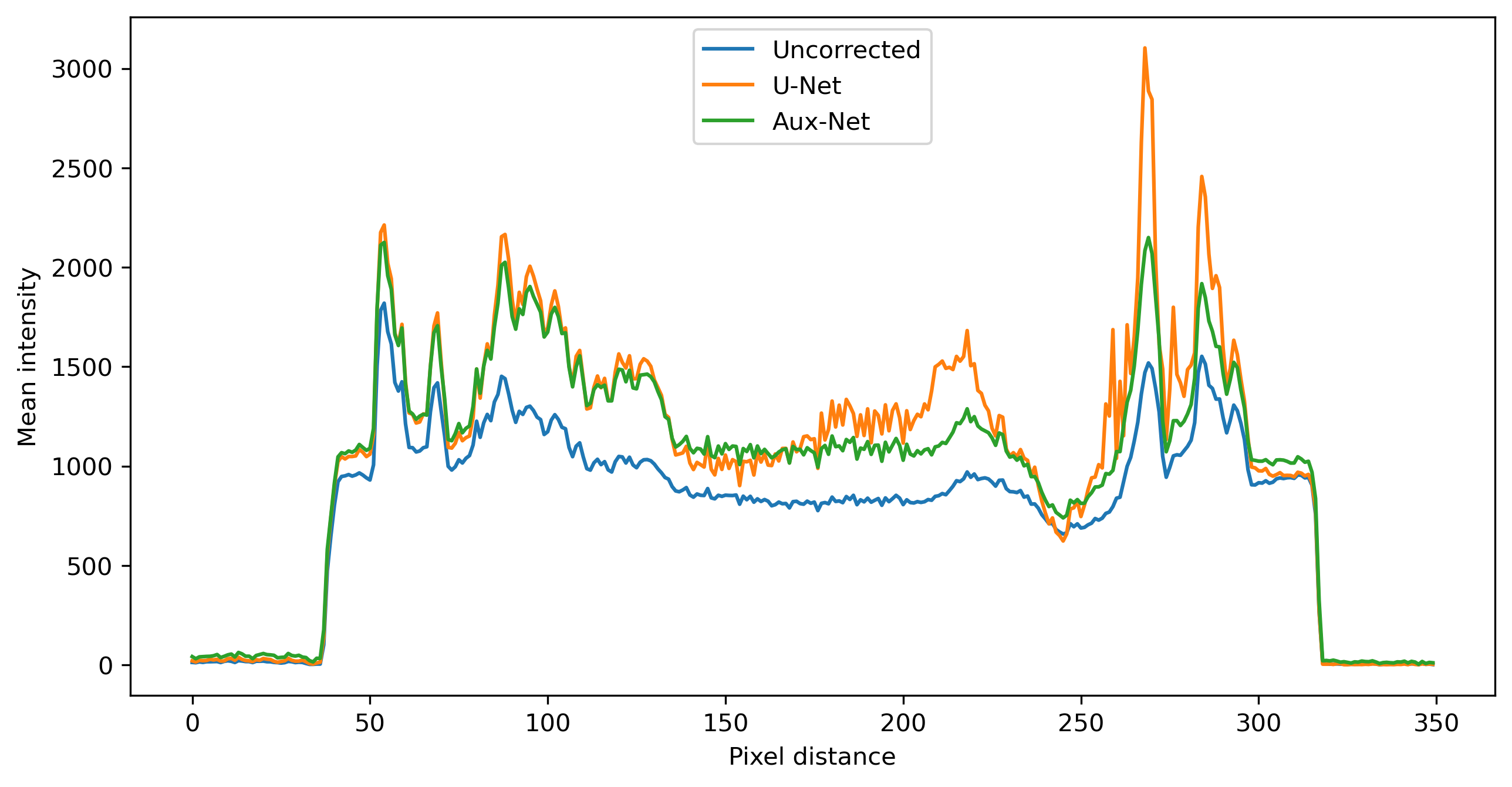}
        \caption{Mean intensity (HU) profiles.}
        \label{fig:subfig4}
    \end{subfigure}

    \begin{subfigure}[b]{0.49\textwidth}
    \centering
        \includegraphics[height=3cm]{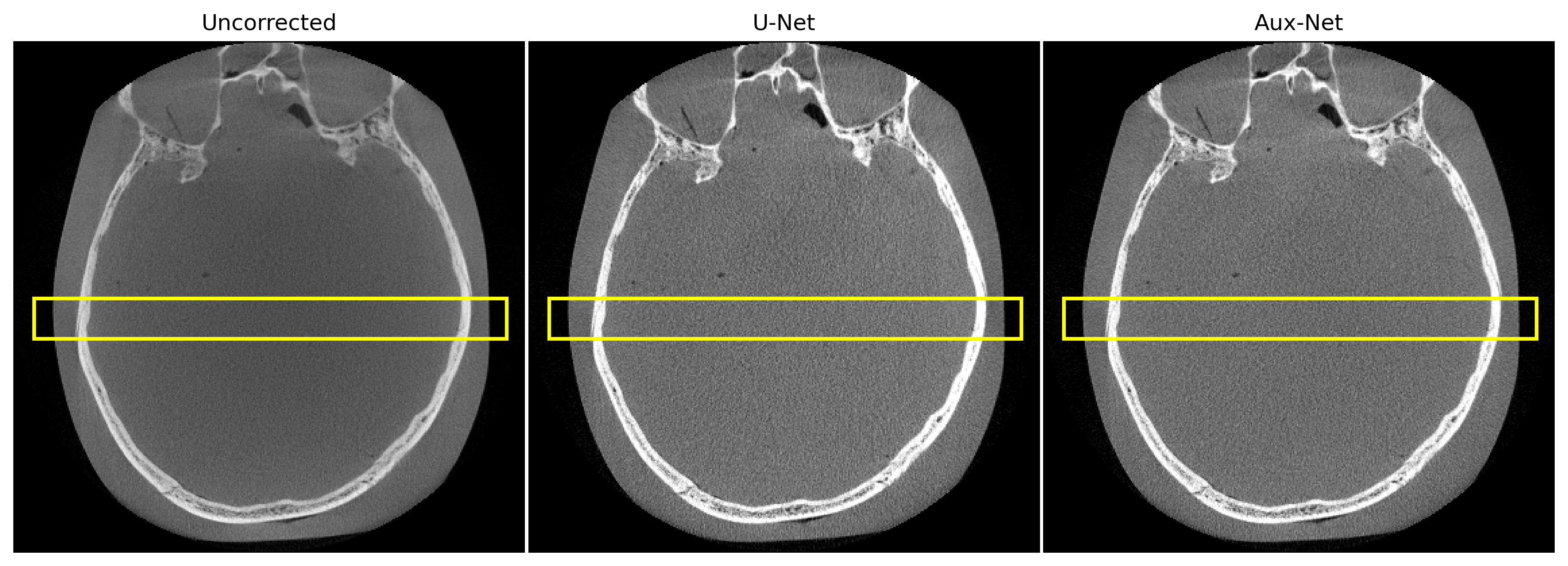}
        \caption{Axial image slice number 221.}
        \label{fig:subfig5}
    \end{subfigure}
    \hfill
    \begin{subfigure}[b]{0.49\textwidth}
    \centering
        \includegraphics[height=3cm]{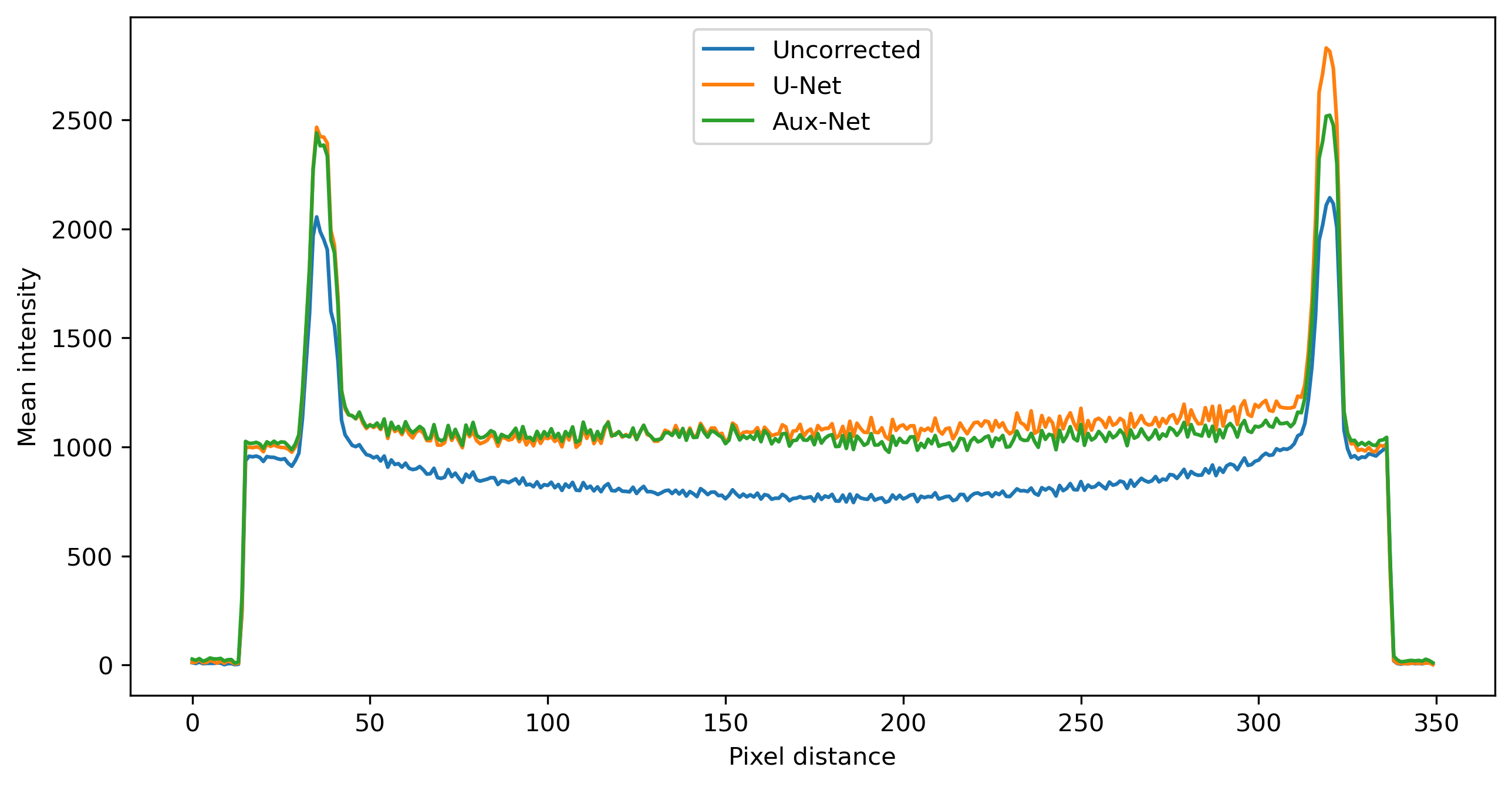}
        \caption{Mean intensity (HU) profiles.}
        \label{fig:subfig6}
    \end{subfigure}\

    \caption{Comparison of the scatter corrected reconstructions from U-Net and Aux-Net with the uncorrected reconstruction of an anthropomorphic phantom. The intensity window of the axial images is -1000 to 2500 HU. Mean intensity profiles of the rectangular area are plotted on the right.}
    \label{fig:images}
\end{figure}

Figure \ref{fig:images} depicts the result of U-Net and Aux-Net based scatter corrected reconstructions for the two axial slices at different locations. The predicted scatter estimates were subtracted from the measurements and reconstruction was done using FDK algorithm for CBCT. The intensity profile is more stable for Aux-Net corrected image slices. Furthermore, the Aux-Net corrected reconstruction is more robust in the presence of high density object in comparison to the U-Net corrected reconstruction.

\section{Conclusion}
In this study, we developed a new approach for improving deep learning based scatter estimation across various FOV sizes. By concatenating the size information with the input of the encoders in the U-Net, we increased the robustness and generalization capability of the scatter estimation model. The proposed Aux-Net outperformed the U-Net for all FOV sizes, specifically on bigger FOVs. Moreover, the proposed method of using auxiliary information is not confined solely to the FOV size. Including other variables that affect the amount of scatter, such as, kV (kilovolt), patient size, object to detector distance, scanned anatomy, and exposure time, as auxiliary information may further enhance the robustness of the neural network and its clinical applicability.

\acknowledgments 
 
The work was supported in-part by a Business Finland grant under TomoHead project.

{
\footnotesize

\bibliography{main} 
\bibliographystyle{spiebib} 
}

\end{document}